\documentclass[preprintnumbers,amsmath,amssymb,floatfix,12pt,prd,superscriptaddress,nofootinbib]{revtex4}
\usepackage{graphicx}
\usepackage{epsfig}
\usepackage{bm}
\usepackage{amsfonts}

\begin{document}
\title{Restoring New Agegraphic Dark Energy in RS II Braneworld}

\author{M. Jamil}
\email{mjamil@camp.nust.edu.pk} \affiliation{Center for Advanced
Mathematics and Physics, National University of Sciences and
Technology, Rawalpindi, 46000, Pakistan}

\author{K. Karami}
\email{KKarami@uok.ac.ir} \affiliation{Department of Physics,
University of Kurdistan, Pasdaran St., Sanandaj, Iran}
\affiliation{Research Institute for Astronomy $\&$ Astrophysics of
Maragha (RIAAM), Maragha, Iran}

\author{A. Sheykhi}
\email{sheykhi@mail.uk.ac.ir} \affiliation{Department of Physics,
Shahid Bahonar University, P.O. Box 76175, Kerman, Iran}
\affiliation{Research Institute for Astronomy $\&$ Astrophysics of
Maragha (RIAAM), Maragha, Iran}

\date{\today}

\begin{abstract}
\vspace*{1.5cm} \centerline{\bf Abstract} \vspace*{1cm} Motivated by
recent works \cite{manos,ahmad}, we investigate new agegraphic model
of dark energy in the framework of RS II braneworld. We also include
the case of variable gravitational constant in our model.
Furthermore, we establish correspondence between the new agegraphic
dark energy with other dark energy candidates based on scalar
fields.
\end{abstract}
\maketitle
\newpage
\section{Introduction}\label{Int}
An interesting attempt for probing the nature of dark energy (DE) is
the so-called ``agegraphic DE'' (ADE). This model was recently
proposed \cite{Cai1} to explain the acceleration of the universe
expansion within the framework of a fundamental theory such as
quantum gravity. The ADE model assumes that the observed DE comes
from the spacetime and matter field fluctuations in the universe.
Following the line of quantum fluctuations of spacetime, Karolyhazy
et al. \cite{Kar1} discussed that the distance $t$ in Minkowski
spacetime cannot be known to a better accuracy than $\delta{t}=\beta
t_{p}^{2/3}t^{1/3}$ where $\beta$ is a dimensionless constant of
order unity. Based on Karolyhazy relation and  Maziashvili arguments
\cite{Maz}, Cai proposed the original ADE model to explain the
acceleration of the universe expansion \cite{Cai1}. Since the
original ADE model suffers from the difficulty to describe the
matter-dominated epoch, a new model of ADE was proposed by Wei and
Cai \cite{Wei2}, while the time scale was chosen to be the conformal
time instead of the age of the universe. The ADE models have arisen
a lot of enthusiasm recently and have examined and studied in ample
detail \cite{age,shey1,shey2,Karami1,Karami2}.

Most researches on the DE puzzle remain in the standard
four-dimensional cosmology. However, in recent years theories of
large extra dimensions, in which the observed universe is realized
as a brane embedded in a higher-dimensional space–time, have
received a lot of interest. According to the braneworld scenario,
the standard model of particle fields is confined to the brane
while, in contrast, the gravity is free to propagate in the whole
space–time \citep{RSII}. In this theory the cosmological evolution
on the brane is described by an effective Friedmann equation that is
incorporated nontrivially with the effects of the bulk into the
brane \citep{Bin}. Apart from being closer to a higher-dimensional
fundamental theory of nature, braneworld has also great
phenomenological successes and a large amount of current research
heads towards this direction. It is therefore desirable to extend
ADE in the braneworld context. The investigations on the DE models
in the context of braneworld scenarios have been carried out in
\cite{Setare0,Setare1}. In the context of ADE, braneworld model with
bulk-brane interaction has also been studied in \cite{ahmad}.

In the present work we would like to restore new ADE (NADE) in RS
II braneworld in the presence of varying gravitational constant.
Employing the agegraphic model of DE in a non-flat universe, we
obtain the equation of state (EoS) parameter for NADE density in
the framework braneworld. We also investigate the correspondence
between NADE and scalar field models of DE such as quintessence,
tachyon, K-essence and dilaton scalar fields and obtain the
evolutionary form of these fields with varying $G$. There are
significant indications that $G$ can by varying, being a function
of time or equivalently of the scale factor \cite{Inn}. In
particular, observations of Hulse-Taylor binary pulsar B$1913+16$
lead to the estimation $\dot{G}/G\sim2\pm4\times10^{-12}{\rm
yr}^{-1}$ \cite{Damour,kogan}, while helio-seismological data
provide the bound $-1.6\times10^{-12}{\rm yr}^{-1}<\dot{G}/G<0$
\cite{guenther}. Similarly,  Type Ia supernova observations  give
the best upper bound of the variation of $G$ as $-10^{-11} {\rm
yr}^{-1} \leq \frac{\dot G}{G}<0$ at redshifts $z \simeq 0.5$
\cite{Gaztanaga}, while astroseismological data from the pulsating
white dwarf star G117-B15A lead to $\left|\frac{\dot G}{G}\right|
\leq 4.10 \times 10^{-11} {\rm yr}^{-1}$ \cite{Biesiada}. In
addition, a varying $G$ has some theoretical advantages too,
alleviating the dark matter problem \cite{Goldman}, the cosmic
coincidence problem \cite{Jamil} and the discrepancies in Hubble
parameter value \cite{Ber}.

This paper is organized as follows. In sections \ref{New} and
\ref{Newi}, we review the formalism of NADE in RS II braneworld.
In section \ref{NewG}, we study the NADE in braneworld with
variable Newton's gravitational constant. In section
\ref{NewField}, we establish the correspondence between NADE model
with other DE candidates based on scalar fields with the
assumption that the gravitational constant $G$ varies with time.
The last section is devoted to conclusions and discussions.

\section{NADE in RS II braneworld}\label{New}
In this section, we apply the bulk NADE in general in RS II
braneworld. The corresponding braneworld action is given by
\cite{manos}
\begin{equation}\label{axon}
S=\int {\rm d}^5x\sqrt{-g}(M_5^3R-\rho_{\Lambda5})+\int{\rm
d}^4x\sqrt{-\gamma}(L_{br}-V+r_cM_5^3R_4).
\end{equation}
Here $M_5$ is the five dimensional Planck mass and $R$ is the
curvature scalar of the five dimensional bulk spacetime. Contrary
to \cite{manos}, we identify $\rho_{\Lambda5}$ as the bulk NADE.
Also $g$ is the determinant of the five dimensional bulk spacetime
metric while $\gamma$ is the four-dimensional spacetime metric.
The quantity $V$ is called the brane tension and $L_{br}$ is the
brane matter content. Also $r_c$ is a characteristic length scale
while $R_4$ is the four dimensional curvature scalar.

The evolution of the brane is given by
\begin{equation}\label{fe}
H^2+\frac{k}{a^2}=\frac{8\pi}{3M_p^2}\rho+\frac{8\pi}{3M_p^2}\rho_\Lambda.
\end{equation}
Here $\rho=\rho_m+\rho_{\Lambda}$,  the sum of energy density of
matter and DE while $M_p^2=(8\pi G)^{-1}$ is the reduced Planck
mass. Following \cite{manos}, the energy density of the four
dimensional effective DE is given by
\begin{equation}\label{rhol}
\rho_{\Lambda}\equiv\rho_{\Lambda4}= \frac{M_p^2}{32\pi
M_5^3}\rho_{\Lambda5}+\frac{3M_p^2}{2\pi\Big( \frac{L_5}{8\pi}-2r_c
\Big)^2}.
\end{equation}
Following \cite{manos}, we have
\begin{equation}\label{rho5}
\rho_{\Lambda5}=c^2\frac{3}{4\pi}M_5^3L^{-2}.
\end{equation}
Thus using (\ref{rho5}) in (\ref{rhol}), we obtain
\begin{equation}\label{rho2}
\rho_{\Lambda}=\frac{3c^2}{128\pi^2}M_p^2L^{-2}+\frac{3M_p^2}{2\pi\Big(
\frac{L_5}{8\pi}-2r_c \Big)^2}.
\end{equation}
In order to use the above expression as the NADE, we replace
$L=\eta$, i.e. the cut-off scale is taken as the conformal age of
the universe defined by
\begin{equation}
\eta=\int_0^a\frac{{\rm d}a}{a^2H}.
\end{equation}
Following \cite{manos}, we are interested in the restored ADE
without bothering about bulk boundaries. Hence we take $L_5$ to be
arbitrary large and ignore the second term in (\ref{rho2}). We
obtain
\begin{equation}
\rho_{\Lambda}=\frac{3c^2}{128\pi^2}M_p^2\eta^{-2}.\label{NADE}
\end{equation}
From (\ref{fe}), we can write
\begin{equation}
1+\Omega_k=\Omega_m+2\Omega_{\Lambda},
\end{equation}
where we have used
\begin{equation}\label{def}
\Omega_k=\frac{k}{(aH)^2},\ \
\Omega_m=\frac{8\pi\rho_m}{3M_p^2H^2},\ \
\Omega_{\Lambda}=\frac{8\pi\rho_{\Lambda}}{3M_p^2H^2}.
\end{equation}
Using (\ref{NADE}) and the last Eq. (\ref{def}), we get
\begin{equation}\label{ol}
\Omega_{\Lambda}=\frac{c^2}{16\pi(H\eta)^2}.
\end{equation}
Differentiating (\ref{NADE}) w.r.t. $t$ and using (\ref{ol}), we
obtain
\begin{equation}\label{drl}
\dot{\rho}_{\Lambda}=-\frac{8H\rho_{\Lambda}}{ac}\sqrt{\pi\Omega_{\Lambda}}.
\end{equation}
Differentiating (\ref{ol}) w.r.t. $t$, we obtain
\begin{equation}\label{orl}
\Omega'_{\Lambda}=-2\Omega_{\Lambda}\Big( \frac{\dot
H}{H^2}+\frac{4}{ac}\sqrt{\pi\Omega_{\Lambda}}\Big),
\end{equation}
where we have used $\dot{\Omega}_{\Lambda}=\Omega'_{\Lambda}H$.
The energy conservation equations are
\begin{equation}
\dot{\rho}_{\Lambda}+4H(1+\omega_{\Lambda})\rho_{\Lambda}=0,\label{r1}
\end{equation}
\begin{equation}
\dot{\rho}_m+4H\rho_m=0.\label{r2}
\end{equation}
Differentiating (\ref{fe}) w.r.t. $t$ and using (\ref{drl}) and
(\ref{r2}) yields
\begin{equation}\label{hdot}
\frac{\dot
H}{H^2}=-2-\Omega_k+4\Omega_{\Lambda}\Big(1-\frac{2}{ac}\sqrt{\pi\Omega_{\Lambda}}\Big).
\end{equation}
Using (\ref{hdot}) in (\ref{orl}), we get
\begin{equation}\label{apl}
\Omega'_{\Lambda}=2\Omega_{\Lambda}\Big[\Omega_k+2(1-2\Omega_{\Lambda})\Big(
1-\frac{2}{ac}\sqrt{\pi\Omega_{\Lambda}}\Big) \Big].
\end{equation}
From (\ref{drl}) and (\ref{r1}), we get
\begin{equation}
\omega_{\Lambda}=-1+\frac{2}{ac}\sqrt{\pi\Omega_{\Lambda}}.
\end{equation}
The deceleration parameter is
\begin{equation}
q=-1-\frac{\dot H}{H^2},
\end{equation}
which yields
\begin{equation}
q=1+\Omega_k-4\Omega_{\Lambda}\Big(1-\frac{2}{ac}\sqrt{\pi\Omega_{\Lambda}}\Big).
\end{equation}

\section{NADE with interaction}\label{Newi}
Assuming an interaction of NADE with matter, the energy conservation
equations take the form
\begin{eqnarray}\label{ece}
\dot{\rho}_{\Lambda}+4H(1+\omega_{\Lambda})\rho_{\Lambda}&=&-Q,\label{r1i}\\
\dot{\rho}_m+4H\rho_m&=&Q,
\end{eqnarray}
where $Q=4b^2H(\rho_m+2\rho_\Lambda)$ is an interaction term, with
$b^2$ is a coupling parameter. Thus from (\ref{drl}) and
(\ref{r1i}), we get
\begin{equation}
\omega_{\Lambda}=-1+\frac{2}{ac}\sqrt{\pi\Omega_{\Lambda}}-\frac{b^2(1+\Omega_k)}{\Omega_\Lambda}.
\end{equation}

\section{NADE with variable Newton's gravitational constant}\label{NewG}
We now treat Newton's gravitational constant to be time dependent
parameter, i.e. $G(t)$. Thus differentiating (\ref{NADE}) w.r.t.
$t$ and using (\ref{ol}), we obtain
\begin{equation}
\dot{\rho}_{\Lambda}=-H\rho_{\Lambda}\Big(\frac{8}{ac}\sqrt{\pi\Omega_{\Lambda}}+
\frac{G'}{G}\Big).\label{dr1g}
\end{equation}
From (\ref{r1}) and (\ref{dr1g}), the EoS parameter becomes
\begin{equation}
\omega_{\Lambda}=-1+\frac{2}{ac}\sqrt{\pi\Omega_{\Lambda}}+\frac{G'}{4G}.\label{wNADE}
\end{equation}
Differentiating (\ref{fe}) w.r.t. $t$, and using (\ref{r2}) and
(\ref{dr1g}) we get
\begin{equation}\label{fg}
\frac{\dot
H}{H^2}=-2-\Omega_{k}+4\Omega_{\Lambda}\Big(1-\frac{2}{ac}\sqrt{\pi\Omega_{\Lambda}}\Big)+\frac{G'}{2G}(1+\Omega_{k}-2\Omega_{\Lambda}).
\end{equation}
Using (\ref{fg}) in (\ref{orl}), we get
\begin{equation}
\Omega'_{\Lambda}=2\Omega_{\Lambda}\Big[\Omega_k+2(1-2\Omega_{\Lambda})\Big(
1-\frac{2}{ac}\sqrt{\pi\Omega_{\Lambda}}\Big)-\frac{G'}{2G}(1+\Omega_{k}-2\Omega_{\Lambda})
\Big].
\end{equation}
The deceleration parameter becomes
\begin{equation}
q=1+\Omega_k-4\Omega_{\Lambda}\Big(1-\frac{2}{ac}\sqrt{\pi\Omega_{\Lambda}}\Big)-\frac{G'}{2G}(1+\Omega_{k}-2\Omega_{\Lambda}).
\end{equation}

\section{Correspondence between NADE with
scalar field DE models }\label{NewField}

Here like \cite{Karami1,Karami3}, we suggest a correspondence
between the NADE model with the quintessence, tachyon, K-essence
and dilaton scalar field models in braneworld cosmology including
varying $G$. To establish this correspondence, we compare the NADE
density (\ref{NADE}) with the corresponding scalar field model
density and also equate the equations of state for this models
with the EoS parameter given by Eq. (\ref{wNADE}).

\subsection{New agegraphic quintessence model}
The energy density and pressure of the quintessence scalar field
$\phi$ are as follows \cite{Copeland}
\begin{equation}
\rho_Q=\frac{1}{2}\dot \phi^2+V(\phi),\label{ro q}
\end{equation}
\begin{equation}
p_Q=\frac{1}{2}\dot \phi^2-V(\phi).\label{p q}
\end{equation}
The EoS parameter for the quintessence scalar field is given by
\begin{equation}
\omega_Q=\frac{p_Q}{\rho_Q}=\frac{\dot \phi^2-2V(\phi)}{\dot
\phi^2+2V(\phi)}.\label{w q}
\end{equation}
Here we establish the correspondence between the NADE scenario and
the quintessence DE model, then equating Eq. (\ref{w q}) with the
EoS parameter of NADE (\ref{wNADE}), $\omega_Q=\omega_\Lambda$,
and also equating Eq. (\ref{ro q}) with (\ref{NADE}),
$\rho_Q=\rho_\Lambda$, we have
\begin{equation} \dot
\phi^2=(1+\omega_\Lambda)\rho_\Lambda,\label{phidot2-2}
\end{equation}
\begin{equation}
V(\phi)=\frac{1}{2}(1-\omega_\Lambda)\rho_\Lambda.\label{Vphi-2}
\end{equation}
Substituting Eqs. (\ref{NADE}) and (\ref{wNADE}) into Eqs.
(\ref{phidot2-2}) and (\ref{Vphi-2}), one can obtain the kinetic
energy term and the quintessence potential energy as follows
\begin{equation}
\dot\phi^2=\frac{3M_P^2H^2\Omega_{\Lambda}}{8\pi}\left(\frac{G'}{4G}+\frac{2\sqrt{\pi
\Omega_{\Lambda}}}{ac}\right) ,\label{fi dot q}
\end{equation}
\begin{equation}
V(\phi)=\frac{3M_P^2H^2\Omega_{\Lambda}}{8\pi}\left(1-\frac{G'}{8G}-\frac{\sqrt{\pi
\Omega_{\Lambda}}}{ac}\right).\label{pot q}
\end{equation}
From Eqs. (\ref{fi dot q}) one can obtain the evolutionary form of
the quintessence scalar field as
\begin{eqnarray}
\phi(a)-\phi(a_0)=\sqrt{\frac{3}{8\pi}}\int_{a_0}^{a}M_p\left[\Omega_{\Lambda}\Big(\frac{G'}{4G}+\frac{2\sqrt{\pi\Omega_{\Lambda}}}{ac}\Big)\right]
^{1/2}\frac{{\rm d}a}{a},
\end{eqnarray}
where $a_0$ is the scale factor at the present time.
\subsection{New agegraphic tachyon model}
The tachyon field was proposed as a source of the DE and inflation.
A rolling tachyon has an interesting EoS whose parameter smoothly
interpolates between $-1$ and $0$ \cite{gibbons}. This discovery
motivated to take DE as the dynamical quantity, i.e. a variable
cosmological constant and model inflation using tachyons. The
tachyon field has become important in string theory through its role
in the Dirac-Born–-Infeld (DBI) action which is used to describe the
D-brane action \cite{Sen}. The effective Lagrangian density of
tachyon matter is given by \cite{Sen}
\begin{equation}
{\mathcal{L}}=-V(\phi)\sqrt{1+\partial_{\mu}\phi
\partial^{\mu}\phi}.
\end{equation}
The energy density and pressure for the tachyon field are as
following \cite{Sen}
\begin{equation}
\rho_{T}=\frac{V(\phi)}{\sqrt{1-\dot{\phi}^{2}}},\label{rhot}
\end{equation}
\begin{equation}
p_{T}=-V(\phi)\sqrt{1-\dot{\phi}^{2}},
\end{equation}
where $V(\phi)$ is the tachyon potential. The EoS parameter for
the tachyon scalar field is obtained as
\begin{equation}
\omega_{T}=\frac{p_{T}}{\rho_{T}}=\dot{\phi}^{2}-1.\label{wt}
\end{equation}
If we establish the correspondence between the NADE and tachyon
DE, then equating Eq. (\ref{wt}) with the EoS parameter of NADE
(\ref{wNADE}), $\omega_T=\omega_\Lambda$, and also equating Eq.
(\ref{rhot}) with (\ref{NADE}), $\rho_T=\rho_\Lambda$, we obtain
\begin{equation}
\dot{\phi}^{2}=\frac{G'}{4G}+\frac{2\sqrt{\pi
\Omega_{\Lambda}}}{ac},\label{fi dotT}
\end{equation}
\begin{eqnarray}
V(\phi)=\frac{3M_P^2H^2\Omega_{\Lambda}}{8\pi}\left(1-\frac{G'}{4G}-\frac{2\sqrt{\pi
\Omega_{\Lambda}}}{ac}\right)^{1/2}.\label{pot T}
\end{eqnarray}
From Eq. (\ref{fi dotT}), one can obtain the evolutionary form of
the tachyon scalar field as
\begin{equation}
\phi(a)-\phi(a_0)=\int_{a_0}^{a}\frac{{\rm
d}a}{Ha}\left(\frac{G'}{4G}+\frac{2\sqrt{\pi
\Omega_{\Lambda}}}{ac}\right)^{1/2}.
\end{equation}
\subsection{New agegraphic K-essence model}

It is also possible to have a situation where the accelerated
expansion arises out of  modifications to the kinetic energy of
the scalar fields. In this context, the K-essence scalar field
model of DE is used to explain the observed late-time acceleration
of the universe. The K-essence is described by a general scalar
field action which is a function of $\phi$ and
$\chi=\dot{\phi}^2/2$, and is given by \cite{Chiba, Picon3}
\begin{equation} S=\int {\rm d} ^{4}x\sqrt{-{\rm
g}}~p(\phi,\chi),
\end{equation}
where $p(\phi,\chi)$ corresponds to a pressure density as
\begin{equation}
p(\phi,\chi)=f(\phi)(-\chi+\chi^{2}),
\end{equation}
and the energy density of the field $\phi$ is
\begin{equation}
\rho(\phi,\chi)=f(\phi)(-\chi+3\chi^{2}).\label{rhok}
\end{equation}
The EoS parameter for the K-essence scalar field is obtained as
\begin{equation}
\omega_{K}=\frac{p(\phi,\chi)}{\rho(\phi,\chi)}=\frac{\chi-1}{3\chi-1}.\label{wk}
\end{equation}
Equating Eq. (\ref{wk}) with the EoS parameter (\ref{wNADE}),
$\omega_{K}=\omega_{\Lambda}$, we find the solution for $\chi$
\begin{equation}
\chi=\frac{1-\frac{G'}{8G}-\frac{\sqrt{\pi
\Omega_{\Lambda}}}{ac}}{2-\frac{3G'}{8G}-\frac{3\sqrt{\pi
\Omega_{\Lambda}}}{ac}}.\label{khi}
\end{equation}
Using $\dot{\phi}^2=2\chi$ and (\ref{khi}), we obtain the
evolutionary form of the K-essence scalar field as
\begin{equation}
\phi(a)-\phi(a_0)=\int_{a_0}^{a}\frac{{\rm
d}a}{Ha}\left(\frac{2-\frac{G'}{4G}-\frac{2\sqrt{\pi
\Omega_{\Lambda}}}{ac}}{2-\frac{3G'}{8G}-\frac{3\sqrt{\pi
\Omega_{\Lambda}}}{ac}}\right)^{1/2}.\label{K F}
\end{equation}
\subsection{New agegraphic dilaton model}
The dilaton scalar field model of DE is obtained from the
low-energy limit of string theory. It is described by a general
four-dimensional effective low-energy string action. The
coefficient of the kinematic term of the dilaton can be negative
in the Einstein frame, which means that the dilaton behaves as a
phantom-type scalar field. However, in presence of higher-order
derivative terms for the dilaton field $\phi$ the stability of the
system is satisfied even when the coefficient of $\dot{\phi}^2$ is
negative \cite{Gasperini}. The pressure (Lagrangian) density and
the energy density of the dilaton DE model is given by
\cite{Gasperini}
\begin{equation}
p_{D}=-\chi+c'e^{\lambda\phi}\chi^{2},
\end{equation}
\begin{equation}
\rho_{D}=-\chi+3c'e^{\lambda\phi}\chi^{2},\label{rhod}
\end{equation}
where $c'$ and $\lambda$ are positive constants and
$\chi=\dot{\phi}^2/2$. The EoS parameter for the dilaton scalar
field is given by
\begin{equation}
\omega_{D}=\frac{p_D}{\rho_D}=\frac{-1+c'e^{\lambda\phi}\chi}{-1+3c'e^{\lambda\phi}\chi}.\label{wd}
\end{equation}
Equating Eq. (\ref{wd}) with the EoS parameter (\ref{wNADE}),
$\omega_D=\omega_\Lambda$, we find the following solution
\begin{equation}
c'e^{\lambda\phi}\chi=\frac{1-\frac{G'}{8G}-\frac{\sqrt{\pi
\Omega_{\Lambda}}}{ac}}{2-\frac{3G'}{8G}-\frac{3\sqrt{\pi
\Omega_{\Lambda}}}{ac}},\label{khi D}
\end{equation}
then using $\dot{\phi}^2=2\chi$, we obtain
\begin{equation}
e^{\frac{\lambda\phi}{2}}\dot{\phi}=\left(\frac{2-\frac{G'}{4G}-\frac{2\sqrt{\pi
\Omega_{\Lambda}}}{ac}}{c'\Big(2-\frac{3G'}{8G}-\frac{3\sqrt{\pi
\Omega_{\Lambda}}}{ac}\Big)}\right)^{1/2}.
\end{equation}
Integrating with respect to $a$, we get
\begin{equation}
e^\frac{\lambda\phi(a)}{2}=e^\frac{\lambda\phi(a_0)}{2}+\frac{\lambda}{2\sqrt{c'}}\int_{a_0}^{a}\frac{{\rm
d}a}{Ha}\left(\frac{2-\frac{G'}{4G}-\frac{2\sqrt{\pi
\Omega_{\Lambda}}}{ac}}{2-\frac{3G'}{8G}-\frac{3\sqrt{\pi
\Omega_{\Lambda}}}{ac}}\right)^{1/2}.
\end{equation}
Therefore the evolutionary form of the dilaton scalar field is
obtained as
\begin{eqnarray}
\phi(a)=\frac{2}{\lambda}\ln\left[e^\frac{\lambda\phi(a_0)}{2}+\frac{\lambda}{\sqrt{2c'}}\int_{a_0}^{a}\frac{{\rm
d}a}{Ha}\left(\frac{1-\frac{G'}{8G}-\frac{\sqrt{\pi
\Omega_{\Lambda}}}{ac}}{2-\frac{3G'}{8G}-\frac{3\sqrt{\pi
\Omega_{\Lambda}}}{ac}}\right)^{1/2}\right].
\end{eqnarray}

\section{Conclusions and discussions}
In this work we studied the NADE model in the framework of RS II
braneworld scenario and restored the EoS as well as the
deceleration parameters in a non-flat universe. Then, we extended
our study to the case where the gravitational constant $G$ varies
with time. A varying $G$ has some theoretical advantages such as
alleviating the dark matter problem \cite{Goldman}, the cosmic
coincidence problem \cite{Jamil} and the discrepancies in Hubble
parameter value \cite{Ber}. We also established a correspondence
between NADE and quintessence, tachyon, K-essence and dilaton
energy density in the braneworld scenario including varying $G$.
We adopted the viewpoint that these scalar field models of DE are
effective theories of an underlying theory of DE. Thus, we should
be capable of using these scalar field models to mimic the
evolving behavior of the NADE and reconstructing the scalar field
models according to the evolutionary behavior of the NADE.
Finally, We reconstructed the potentials and the dynamics of these
scalar field models which describe quintessence, tachyon,
K-essence and dilaton cosmology.

\begin{acknowledgments}
The works of K. Karami and A. Sheykhi have been supported
financially by Research Institute for Astronomy $\&$ Astrophysics
of Maragha (RIAAM), Maragha, Iran.
\end{acknowledgments}

\end{document}